# Properties and Behaviors of Heavy Quarkonia: Insights Through Fractional Model and Topological Defects

M. Abu-shady[1] and H. M. Fath-Allah[2]

Department of Mathematics and Computer Sciences, Faculty of Science, Menoufia University, Egypt[1]

Higher Institute of Engineering and Technology, Menoufia, Egypt[2]

## Abstract

In this study, we investigated the impact of a topological defect ($\lambda$) on the properties of heavy quarkonia using the extended Cornell potential. We solved the fractional radial Schrödinger equation (SE) using the extended Nikorov-Uvarov (ENU) method to obtain the eigen energy, which allowed us to calculate the masses of charmonium and bottomonium. One significant observation was the splitting between np and nd states, which we attributed to the presence of the topological defect. We discovered that the excited states were divided into components corresponding to $2l + 1$, indicating that the gravity field induced by the topological defect interacts with energy levels in a manner similar to the Zeeman effect caused by a magnetic field. Additionally, we derived the wave function and calculated the root mean radii for charmonium and bottomonium. A comparison with classical models was performed, resulting in better results being obtained. Furthermore, we investigated the thermodynamic properties of charmonium and bottomonium, determining quantities such as energy, partition function, free energy, mean energy, and specific heat for p-states. The obtained results were found to be consistent with experimental data and previous works. In conclusion, the fractional model used in this work proved essential in understanding the various properties and behaviors of heavy quarkonia in the presence of topological defects.

## 1. Introduction

The thorough description of hadrons properties has become an important and significant issue in particle and nuclear physics. There have been many attempts to improve chiral quark models in order to calculate hadron properties, such as [1, 2, 3, 4, 5]. Additionally, these models have been extended to quark-gluon plasma in hot or dense mediums, as discussed in [6, 7, 8].

By the end of the 1970s, in the existence of Grand Unification Theories (GUT), Kibble predicted that a sequence of spontaneous symmetry breaks would occur in our universe during the cooling phase after the Big Bang [9-10]. Phase transitions were present in conjunction with these symmetry breaks. According to the Kibble-introduced mechanism, these phase transitions would typically have produced some topological flaws. These flaws are areas of space where energy is concentrated at a very high density. Their spatial dimensions define their natures. These flaws produce the GUT's structure of the collection of empty spaces that the Higgs fields can access. Numerous topological defects, including point defects known as monopoles [9, 11], linear defects or cosmic strings [9, 10, 12], surface defects or domain walls, and combinations of these defects have been examined in the past, surface defects, domain walls, and combinations of these defects, among others. Since topological flaws are frequently persistent, it is quite likely that some of them have persisted, possibly even up to the present. Cosmology can detect topological flaws. The dynamics of our universe would be swiftly dominated by monopoles and domain barriers since they are so enormous, which is completely at odds with the observations. Cosmological consequences of cosmic strings could be proven by present and future research, and they are consistent with current observations. The geometry of cosmic strings is flat everywhere except for the symmetry axis, making them the most notable topological defects [9, 11, 13]. They were created early in the universe's history. The variety of theories produced from General Relativity Theory is a very strong incentive to investigate how particles behave on these geometrical structures, despite the lack of any fundamental evidence for their existence. The importance of quantum systems on space times with linear defect geometry should, therefore, receive special attention, as they are thought to constitute the most significant topological flaw in our universe [12]. We can list, for instance, the compression of matter during a moving string's passage and temperature variations in the Cosmic Microwave Background (CMB) as key effects of the hypothetical presence of cosmic strings. A cosmic string's gravitational field has peculiar characteristics. In fact, a quantum particle at rest near a static, straight line with infinite dimensions wouldn't be drawn to it and wouldn't experience any local gravitational fields as a result. This indicates that, close to a cosmic string, space is flat locally [11]. It is not a worldwide flatness, though; it is local. A cosmic string can have a range of effects on quantum particle dynamics because of its particular shape, including the generation or destruction of the (e+, e-) pair [14] and the bremsstrahlung process [15] in the vicinity of a static nucleus. We become interested in the characteristics of hadrons in such spaces under the influence of a central field as a result of these cosmic string-induced fluctuations of quantum observables.

Theoretical high-energy physics (HEP) and related sciences heavily rely on the study of the thermodynamic features of quantum systems [16]. Quarks and gluons typically stay contained in hadrons, especially in the protons and neutrons that make up the atomic nucleus. In fact, a plasma of quarks and gluons with poorly understood thermodynamic properties can be produced by compressing or heating nuclei. A theory that explains the strong interaction is called quantum chromodynamics [17, 18]. Several experimental studies have been conducted in recent years to discover the presence of deconfinement transitions [18]. A bound state of charm and anti-charm quarks called the $c\bar{c}$, which is suppressed, was then suggested as a potential indicator of the QCD phase change [17, 18]. In-depth research on the thermodynamic characteristics in the case of heavy quarkonia would be intriguing, given that the suppression of the $c\bar{c}$ meson has been identified as the signature of the deconfinement transition. In [19], the author employed the quark-gluon plasma component quasi-particle model to derive the thermodynamic parameters of the system. Then, utilizing chiral quark models, thermodynamic properties are examined in [20, 21].

In this paper, we derive the SE solutions caused by the gravitational field of a $\lambda$ for all values of the orbital angular momentum quantum number $l_1$ . The radial Schrödinger equation in the cosmic-string background is solved analytically using the generalized fractional extended NU method. Then, we obtain the results to calculate the mass of charmonium and bottomonium, the root mean and the thermodynamic properties of heavy quarkonia from cosmic-string geometry which are not considered in previous works in the framework of fractional non-relativistic quark models.

This paper is organized as follows: Section 2 reviews the generalized fractional derivative of the ENU method. We obtain the solution for the SE in the cosmic-string background for the extended Cornel potential with generalized fraction derivative of extended NU method in Section 3. Results are the special cases, the mass, root mean and thermodynamic properties of charmonium and bottomonium inside linear defect geometry are then described in section 4. Conclusions are provided in Section 5

## 2. The Generalized Fractional Derivative of the Extended Nikiforvo-Uvarov Method with (GFD-ENU).

ENU method is a generalization of the Nikiforv-Uvarov method. Both are frequently used in quantum physics to determine the eigenvalue and eigenfunctions of the Schrodinger or Dirac equations, as well as any other equations that need to be transformed into a hypergeometric form for a full analysis, for further information, see Refs. [22, 23]. In order to demonstrate the method viability, the NU was successfully applied in a few practical examples in Ref. [24], where it was generalized to conformable fractional derivative. The goal of this section is to expand the ENU within the confines of the GFD. Take into account the generalized fractional differential equations shown in standard from below [25]:

$$D^\alpha [D^\alpha[\Psi(x)]] + \frac{\tilde{\tau}(x)}{\sigma(s)} D^\alpha[\Psi(x)] + \frac{\tilde{\sigma}(x)}{\sigma^2(x)} \Psi(x) = 0, \qquad (1)$$

where $\tilde{\tau}(x), \sigma(x)$ and $\tilde{\sigma}(x)$ are polynomials with degrees of no more than second, third and fourth, respectively. Then using GFD [26], we can write

$$D^\alpha[\Psi(x)] = \frac{\Gamma(\beta)}{\Gamma(\beta-\alpha-1)} x^{1-\alpha} \Psi´(x), \qquad (2)$$

$$D^\alpha [D^\alpha[\Psi(x)]] = \left(\frac{\Gamma(\beta)}{\Gamma(\beta-\alpha-1)}\right)^2 [(1-\alpha) x^{1-2\alpha}\Psi´(x) + x^{2-2\alpha} \Psi''(x)] \qquad (3)$$

where $0 < \alpha, \beta \leq 1$, from of Eqs. 2,3 we get,

$$\Psi''(s) + \frac{\tilde{\tau}(x) + (1-\alpha) x^{1-\alpha}\sigma(x)}{x^{1-\alpha} \sigma(x)}\Psi´(x) + \left(\frac{\Gamma(\beta)}{\Gamma(\beta-\alpha-1)}\right)^{-2} \frac{\breve{\sigma}(x)}{s^{2-2\alpha} \sigma^2(x)} \Psi(x) = 0 \qquad (4)$$

By comparing the Eqs. (1, 4), we get the fractional parameters,

$$\tilde{\tau}_f(x) = \tilde{\tau}(x) + (1-\alpha) x^{1-\alpha}\sigma(x),$$

$$\sigma_f(x) = x^{1-\alpha} \sigma(x), \qquad (5)$$

$$\tilde{\sigma}_f(x) = \left(\frac{\Gamma(\beta)}{\Gamma(\beta-\alpha-1)}\right)^{-2} \tilde{\sigma}(x),$$

then we obtain the standard equation (GFD-ENU)

$$\Psi''(x) + \frac{\tilde{\tau}_f(x)}{\sigma_f(x)} \Psi'(x) + \frac{\tilde{\sigma}_f(x)}{\sigma_f^2(x)} \Psi(x) = 0. \qquad (6)$$

We assume that the following transformation will provide the answer to Equ. (1)

$$\Psi(x) = \Phi(x)\, Y(x), \qquad (7)$$

Eq. (6) is reduced to a hypergeometric equation then we get,

$$\sigma_f(x)\, Y''(x) + \tau_f(x)\, Y'(x) + \lambda_f(x)\, Y(x) = 0, \qquad (8)$$

where $\Phi(s)$ adchive

$$\frac{\Phi'(x)}{\Phi(x)} = \frac{\Pi_f(x)}{\sigma_f(x)} \qquad (9)$$

$$\lambda_f(x) - \Pi_f'(x) = G(x) \qquad (10)$$

$Y(x)$ is the hypergeometric function has a polynomial that archive the Rodrigues relation

$$Y_n(s) = \frac{B_n}{\rho(x)} \frac{d^n}{ds^n}\left(\sigma^n{}_f(x)\, \rho(x)\right), \qquad (11)$$

where $B_n$ is the constant for normalization, and $\rho(x)$ is the weight function and satisfy the relation

$$\sigma'_f\, \rho + \rho'\sigma_f = \tau_f\, \rho, \qquad (12)$$

the function $\Pi_f(s)$ are defined as

$$\Pi_f(x) = \frac{\sigma'_f(x) - \tilde{\tau}_f(x)}{2} \pm \sqrt{\left(\frac{\sigma'_f(x) - \tilde{\tau}_f(x)}{2}\right)^2 - \tilde{\sigma}_f(x) + G(x)\sigma_f(x)}. \qquad (13)$$

$\Pi_f(x)$ is a polynomial of at most $2\alpha$ degree, and based on this the determination of $G(x)$ is important in obtaining $\Pi_f(x)$. $\lambda_n(x)$ is defined from relation

$$\lambda_n(x) = \frac{-n}{2}\tau'_f(x) - \frac{n(n-1)}{6}\sigma_f''(x) \qquad (14)$$

where,

$$\tau_f(x) = \tilde{\tau}_f(x) + 2\,\Pi_f(x) \qquad (15)$$

we obtain the eigen energy from Eq. (10) with Eq. (14).

## 3. Cosmic String Space-Time of Heavy Quarkonia

In spherical coordinates, the line element which shows the linear defect space-time [10] is obtained by ($x^0 = ct$, $x^1 = r$, $x^2 = \theta$, $x^3 = \varphi$)

$$ds^2 = \Omega_{\mu\nu} \, dx^\mu \otimes dx^\nu = -c^2 \, dt^2 + dr^2 + r^2 \, d\theta^2 + [\chi \, d\theta + \lambda \, r \, sin\theta \, d\varphi]^2 \tag{16}$$

where $0 < r < \infty$, $0 < \theta < \pi$, and $0 < \varphi < 2\pi$, $0 < \lambda = 1 - 4T$ is the topological parameter of the cosmic string, $\chi = \frac{4 G T}{c^3}$, is the torsion [27] parameter and T represents the cosmic string's linear mass density. From General Relativity (GR), we note that the values of T varies in the interval $T \in\, ]0,1[$ [27,28].

For $\lambda \to 1$ and $\chi \to 0$, the metric obtained by Equ. (16) reduces to the usual Minkowski metric in spherical coordinates [29, 30].

The metric tensor for the space-time given by Equ. (16) is:

$$\Omega_{\mu\nu}(x) = \begin{bmatrix} -1 & 0 & 0 & 0 \\ 0 & 1 & 0 & 0 \\ 0 & 0 & \chi^2 + r^2 & \chi\lambda r \, sin\theta \\ 0 & 0 & \chi\lambda r \, sin\theta & \lambda^2 r^2 \, sin^2\theta \end{bmatrix} = \begin{bmatrix} -1 & 0 \\ 0 & (\Omega_{ij}) \end{bmatrix} \tag{17}$$

With the inverse metric,

$$\Omega^{\mu\nu}(x) = \begin{bmatrix} -1 & 0 & 0 & 0 \\ 0 & 1 & 0 & 0 \\ 0 & 0 & \frac{1}{r^2} & \frac{-\chi}{\lambda r^3 sin\theta} \\ 0 & 0 & \frac{-\chi}{\lambda r^3 sin\theta} & \frac{\chi^2 + r^2}{\lambda^2 r^4 sin^2\theta} \end{bmatrix} \tag{18}$$

We choice the signature $(-,+,+,+)$ for the metric tensor $\Omega^{\mu\nu}$, and its determinant is obtained by $\Omega = \det(\Omega^{\mu\nu}) = -\lambda^2 r^4 \, sin^2\theta$, with $\mu, \nu = 0,1,2,3$. In the system of curvilinear coordinates $ds^2 = \sum_{i=1}^{3}\sum_{j=1}^{3} \Omega_{ij} \, dx^i \otimes dx^j$ such that $r \to x^1$, $\theta \to x^2$, $\varphi \to x^3$ the 3-dimensional interior Euclidian space's metric tensor is:

$$\Omega_{ij}(x) = \begin{bmatrix} 1 & 0 & 0 \\ 0 & \chi^2 + r^2 & \chi\lambda r \, sin\theta \\ 0 & \chi\lambda r \, sin\theta & \lambda^2 r^2 \, sin^2\theta \end{bmatrix} \tag{19}$$

The Laplace- Beltrami (LB) operator of the local coordinates system may be described as:

$$\Delta_{LB} = \frac{1}{\sqrt{\Omega}} \frac{\partial}{\partial x^i} (\Omega^{ij} \sqrt{\Omega} \frac{\partial}{\partial x^i}) \qquad i,j = 1,2,3 \text{ and } \Omega = \det(\Omega_{ij}) = \lambda^2 r^4 \sin^2\theta \quad (20)$$

Next, considering Equation (20) and for small values of the torsion parameter $\chi \ll 1$, the (LB) operator is:

$$\Delta_{LB} = \frac{1}{r^2} \left\{ \frac{\partial}{\partial r}\left[r^2\left(\frac{\partial}{\partial r}\right)\right] + \cot\theta \frac{\partial}{\partial \theta} + \frac{\partial^2}{\partial \theta^2} + \frac{1}{\lambda^2 \sin^2\theta} \frac{\partial^2}{\partial \varphi^2} \right\}, \quad (21)$$

The Hamiltonian operator in natural can be written from this. ($\hbar = c = 1$) as:

$$H = -\frac{1}{2M}\left[\frac{\partial^2}{\partial r^2} + \frac{2}{r}\frac{\partial}{\partial r} + \frac{1}{r^2}\cot\theta \frac{\partial}{\partial \theta} + \frac{1}{r^2}\frac{\partial^2}{\partial \theta^2} + \frac{1}{r^2}\frac{1}{\lambda^2 \sin^2\theta}\frac{\partial^2}{\partial \varphi^2}\right] + V(r,\theta,\varphi) \quad (22)$$

$M = \frac{m_q m_{\bar{q}}}{m_q + m_{\bar{q}}}$ is the reduced mass, where $m_q, m_{\bar{q}}$ are mass of quark and antiquark [24, 31].

Now, in curved cosmic string space-time, the non-relativistic radial (SE) is presented. detail [12, 27]

$$\frac{d^2\psi_{nl}(r)}{dr^2} + \left[-\frac{2M}{\hbar^2}V(r) + \frac{2M}{\hbar^2}E_{nl} - \frac{\delta}{r^2}\right]\psi_{nl}(r) = 0 \quad (23)$$

Where $\delta = l_{(\lambda)}(l_{(\lambda)} + 1)$ with $l_{(\lambda)} = m_{(\lambda)} + n$ and the quantum number for generalized angular orbits is $l_{(\lambda)}$. It's not necessarily the case that the generalized quantum numbers $l_{(\lambda)}$ and $m_{(\lambda)}$ are integers.. $l_{(\lambda)} = m_{(\lambda)} + n = \frac{m}{\lambda} + n = l_1 - \left(1 - \frac{1}{\lambda}\right)m$, where $l_1 = 0,1,2,\ldots$

Suppose two heavy quarks in a bound state like the $c\bar{c}$ and $b\bar{b}$ with the potential consists of Cornell potential plus harmonic potential [32]

$$V(r) = a r^2 + b r - \frac{c}{r} + d. \quad (24)$$

To put Equ. (23) in the dimensional fractional form, we let $r = \frac{y}{A}$, $M' = \frac{M}{A}$ and $E' = \frac{E}{A}$

where A= 1 GeV. $\delta$ is separation constant $\delta = l_{(\lambda)}(l_{(\lambda)} + 1)$, then we get,

$$D^\alpha [D^\alpha \psi_{nl}(y)] + 2\mu'(E' - \frac{V(y)}{A}) - \frac{\delta}{z^{2\alpha}}\psi_{nl}(y) = 0 \quad (25)$$

where $a' = \frac{a}{A^3}$, $b' = \frac{b}{A^2}$, $c' = c$, $d' = \frac{d}{A}$, $\zeta = \frac{\Gamma(\beta)}{\Gamma(\beta - \alpha - 1)}$.

By using Equ. (3), we get,

$$\psi''_{nl}(y) + \frac{1-\alpha}{z}\psi'_{nl}(y) + \frac{1}{z^2}((\varepsilon - d^1)y^{2\alpha} - a^1 y^{4\alpha} - b^1 y^{3\alpha} + c^1 y^\alpha - \delta^1)\psi_{nl}(y) = 0, \quad (26)$$

where $\varepsilon = \frac{2M'E'}{\zeta^2}$, $a^1 = \frac{2M'a'}{\zeta^2}$, $b^1 = \frac{2M'b'}{\zeta^2}$, $c^1 = \frac{2M'c'}{\zeta^2}$, $d^1 = \frac{2M'd'}{\zeta^2}$ and $\delta^1 = \frac{\delta}{\zeta^2}$

$$\psi''_{nl}(y) + \frac{1-\alpha}{z}\psi'_{nl}(y) + \frac{1}{z^2}(-\varsigma_1 y^{4\alpha} - \varsigma_2 y^{3\alpha} - \varsigma_3 y^{2\alpha} - \varsigma_4 y^\alpha - \varsigma_5)\psi_{nl}(y) = 0, \quad (27)$$

where $\varsigma_1 = a^1$, $\varsigma_2 = b^1$, $\varsigma_3 = -\varepsilon + d^1$, $\varsigma_4 = -c^1$ and $\varsigma_5 = \delta^1$

we get by comparing Equ. (6) and Equ. (27),

$$\tilde{\tau} = 1 - \alpha, \sigma_f = y, \tilde{\sigma}_f = -\varsigma_1 y^{4\alpha} - \varsigma_2 y^{3\alpha} - \varsigma_3 y^{2\alpha} - \varsigma_4 y^\alpha - \varsigma_5 \quad (28)$$

Then we get

$$\Pi_f(y) = \frac{\alpha}{2} \pm \sqrt{\varsigma_1 y^{4\alpha} + \varsigma_2 y^{3\alpha} + \varsigma_3 y^{2\alpha} + \varsigma_4 y^\alpha + \bar{\varsigma}_5 + y\, G(y)}, \quad (29)$$

where $\bar{\varsigma}_5 = \varsigma_5 + \frac{\alpha^2}{4}$, we choice a linear function $G(y) = A^* y^{2\alpha-1} + B\, y^{\alpha-1}$, that produces the functions under the root in above equation be quadratic $(A_1 y^{2\alpha} + A_2 y^\alpha + A_3)^2$. Then

$$\Pi_f(y) = \frac{\alpha}{2} \pm (A_1 y^{2\alpha} + A_2 y^\alpha + A_3) \quad (30)$$

Comparing Eq. (29) and Eq. (30), we obtain

$A_1 = \pm\sqrt{\varsigma_1}$,

$A_2 = \pm\frac{\varsigma_2}{2\sqrt{\varsigma_1}}$,

$A_3 = \pm\sqrt{\bar{\varsigma}_5}$, \quad (31)

$A^* = \frac{\varsigma_2^2}{4\varsigma_1} \pm 2\sqrt{\varsigma_1 \bar{\varsigma}_5} - \varsigma_3$,

$B = \pm\frac{\varsigma_2}{\sqrt{\varsigma_1 \bar{\varsigma}_5}} - \varsigma_4$

$$\tau_f(y) = 1 \pm 2(A_1 y^{2\alpha} + A_2 y^\alpha + A_3), \quad (32)$$

From Eq. (10) we obtain

$$\lambda(y) = A^* y^{2\alpha-1} + B\, y^{\alpha-1} \pm (2 A_1 y^{2\alpha-1} + \alpha A_2 y^{\alpha-1}) \quad (33)$$

From Equ. (14), we obtain

$$\lambda_n(y) = -n(\pm 2 A_1 y^{2\alpha-1} + \alpha A_2 y^{\alpha-1}) \tag{34}$$

We obtain the energy eigenvalue from Eq. (33) and Eq. (34)

$$E_{nlm} = d - \frac{b^2}{4a} + \zeta^2 \sqrt{\frac{2a}{M}} \left[ (n+1)\alpha + \sqrt{\frac{\alpha^2}{4} + \frac{1}{\zeta^2}\left[l_1 - \left(1-\frac{1}{\lambda}\right)m\right]\left[l_1 - \left(1-\frac{1}{\lambda}\right)m + 1\right]} \right] \tag{35}$$

From Equ. (9) we obtain the function $\Phi(y)$

$$\Phi(y) = k\, y^{\frac{\alpha}{2}+A_3}\, e^{\frac{1}{2}\left(\frac{A_1}{\alpha}y^{2\alpha} + \frac{2A_2}{\alpha}y^{\alpha-1}\right)}, \tag{36}$$

where $A_1, A_2$, and $A_3$ are obtained from Equ. (31). From Equ. (12) we obtain the function $\rho(z)$

$$\rho(y) = y^{2A_3} e^{\left(\frac{A_1}{\alpha}y^{2\alpha} + \frac{2A_2}{\alpha}y^{\alpha-1}\right)}, \tag{37}$$

Then the function $Y_n(s)$

$$Y_n(s) = B_n\, y^{-2A_3}\, e^{-\left(\frac{A_1}{\alpha}y^{2\alpha} + \frac{2A_2}{\alpha}y^\alpha\right)} \frac{d^n}{dz^n}\left[y^{2A_3+n}\, e^{\left(\frac{A_1}{\alpha}y^{2\alpha} + \frac{2A_2}{\alpha}y^{\alpha-1}\right)}\right] \tag{38}$$

From Equ. (7) we obtain the fractional radial wave function

$$\psi_{nl}(y) = N_{nl}\, y^{\frac{\alpha}{2}-A_3}\, e^{-\left(\frac{A_1}{2\alpha}y^{2\alpha} + \frac{A_2}{\alpha}y^{\alpha-1}\right)} \frac{d^n}{dy^n}\left[y^{2A_3+n}\, e^{\left(\frac{A_1}{\alpha}y^{2\alpha} + \frac{2A_2}{\alpha}y^{\alpha-1}\right)}\right] \tag{39}$$

where $N_{nl}$ is normalization constant.

## 4. Results and discussion

### 4.1 Special cases

The classical case is obtained ($\alpha = \beta = 1$) then $\zeta = 1$. The eigen energy and wave function in classical case are:

$$E_{nlm} = d - \frac{b^2}{4a} + \sqrt{\frac{2a}{M}} \left[ (n+1) + \sqrt{\frac{1}{4} + \left[l_1 - \left(1-\frac{1}{\lambda}\right)m\right]\left[l_1 - \left(1-\frac{1}{\lambda}\right)m + 1\right]} \right] \tag{40}$$

$$\psi_{nl}(y) = N_{nl}\, y^{\frac{1}{2}-A_3}\, e^{-\left(\frac{A_1}{2}y^2 + A_2\right)} \frac{d^n}{dy^n}\left[y^{2A_3+n}\, e^{(A_1 y^2 + 2A_2)}\right] \tag{41}$$

### 4.2. Mass of heavy quarkonia

We obtain the mass of both charmonium and bottomonium from relation

$$M = m_1 + m_2 + E_{nlm}$$

$$= m_1 + m_2 + d - \frac{b^2}{4a} + \zeta^2 \sqrt{\frac{2a}{M}} \left[ (n+1)\alpha + \sqrt{\frac{\alpha^2}{4} + \frac{1}{\zeta^2} \left[ l - \left(1 - \frac{1}{\lambda}\right)m \right] \left[ l - \left(1 - \frac{1}{\lambda}\right)m + 1 \right]} \right] \quad (42)$$

Table (1), the mass of $c\bar{c}$ have been calculated for 1S, 2S, 3S, 4S, 1P, 2P and 1D for different values of topological defect $\lambda$. We have been calculated the Schrödinger equation by using generalized fraction (ENU) method and we get the mass of $c\bar{c}$ under effect $\lambda$. The potential parameter $a, b$ and d are fitted using Eq. (42) where, $a = 0.0826\ GeV^3$, $b = 0.41849\ GeV^2$ and $d = 0.427269\ GeV$ and the quark mass ($m_c = 1.207\ GeV$) are obtained from Refs. [24, 33]. We obtained the mass of $c\bar{c}$ in two cases the first case was the classical case at ($\alpha = \beta = 1$) at topological defect ($\lambda = 0.8$), we get a good results compared with previous works and some of state of charmonium is close with experimental and we calculated total error in the first case is smaller than previous works is 0.018% but in second case total error is smaller than first case is 0.016%. In the present work topological defect takes values $0 < \lambda < 1$. The topological produces a splitting in the mass spectrum of the $c\bar{c}$ meson in order to break the degeneracy in the nP and nD states, which was absent in earlier studies. In Ref. [34], they studied the N-radial Schrödinger equation by using asymptotic iteration method for the quark-antiquark interaction potential and they calculated the mass spectra of heavy quarkonia so that we calculated total error of this work is 0.9513%. In Ref. [35], they sloved the Schrödinger equation by using Nikiforov-Uavorv method for the sum of a harmonic, a linear and a Coulomb interaction potential and they calculated the mass spectra of heavy quarkonia so that we calculated total error of this research is 0.11065%. In Ref. [36], they calculated the N- radial Schrödinger equation by using the power series iteration method for the quark-antiquark interaction potential and they calculated the mass spectra of heavy quarkonia so that we calculated total error of this research is 0.11152%. In Ref. [37], they studied the N- radial Schrödinger equation by using an exact-analytically iteration method for Trigonometric Rosen-Morse of the quark-antiquark interaction potential and they calculated the mass spectra of heavy quarkonia so that we calculated total error of this research is 0.05788%.

Table (2), the mass of $b\bar{b}$ have been calculated for 1S, 2S, 3S, 4S, 1P, 2P and 1D for different values of topological defect $\lambda$. We have been calculated the Schrödinger equation by using generalized fraction ENU method and we get the mass of $b\bar{b}$ under effect $\lambda$. The potential parameter $a, b$ and d are fitted using Eq.(42), where $a = 0.17776\ GeV^3$, $b = 0.6898\ GeV^2$ and $d = -0.0927487\ GeV$ and the quark mass ($m_b = 4.823\ GeV$) are obtained from Refs. [24, 33]. We obtained the mass of $b\bar{b}$ in two cases the first case was the classical case at ($\alpha = \beta = 1$) at topological defect($\lambda = 0.7$), we get a good results compared with previous works and some of state of Bottomonium is close with experimental and we calculated total error in the first case is smaller than previous works is 0.006% but in second case ($\alpha = 0.8, \beta = 0.9$) at $\lambda = 0.8$) total error is smaller than first case is 0.0055% . The topological produces a splitting in the mass spectrum of the $b\bar{b}$ meson in order to break the degeneracy in the nP and nD states, which was absent in earlier studies. In Ref. [34], they studied the N- radial Schrödinger equation by using asymptotic iteration method for the quark-antiquark interaction potential and they calculated the mass spectra of heavy quarkonia so that we calculated total error of this research is 0.484%. In Ref. [36], they calculated the N-radial Schrödinger equation by using the power series iteration method for the quark-antiquark interaction potential and they calculated the mass spectra of heavy quarkonia so that we calculated total error of this research is 0.0137%. In Ref. [38], they studied the N-radial Schrödinger equation by using asymptotic iteration method with Cornell potential so that we calculated total error of this research is 0.028%.

In Fig. (1), The levels are different from the Minkowski levels, as we can see with $m = 1$ of P-states of charomonium. In left panel, we note that the splitting increase as $\lambda$ decreases compared to right panel because we take $\lambda = 0.4$ in the left panel but right panel is $\lambda = 0.8$. This is an agreement with Ref. [39]. And, we note that different values of $\alpha, \beta < 1$ helps also on appearing the splitting. Since $\lambda < 1$, the degenerated levels appears. In Fig. (2), we note that when $m = 0$ of P-states of charomonium. The topological defect does not effect on the system and there are not splitting is appeared. This is an agreement with Ref. [39], but the generalized fractional parameter plays role where in left panel we take $\alpha = \beta = 0.9$ and in the right panel $\alpha = \beta = 0.4$, we note that in the right panel the curve is higher than the curve in the left panel. In Fig. (3), we observe that of P-states of charmonium when $m = -1$. Since complex eigenenergy appears by the negative sign of $m_l$, some states do not see. This is an agreement with Ref. [39]. Also, the generalized fractional prameter plays role where in the left panel, we take $\alpha = \beta = 0.9$ and in the right panel $\alpha = \beta = 0.4$, we note that in the right panel the curve is higher than the curve in the left panel. In Fig. (4), The levels are different from the Minkowski levels, as we can see with $m = 1$ of P-states of bottomonium. In the left panel, we note that the splitting increase as $\lambda$ decreases compared to the right panel because we take $\lambda = 0.4$ in the left panel but the right panel is $\lambda = 0.8$. This is an agreement with Ref. [39]. And, we note that different values of $\alpha, \beta < 1$ helps also on appearing the splitting. Since $\lambda < 1$, the degenerated levels appears. In Fig. (5), we note that when $m = 0$ of P-states of bottomonium. The topological defect does not effect on the system and there are not splitting is appeared. This is an agreement with Ref. [38], but the generalized fractional parameter plays role where in the left panel we take α = β = 0.9 and in the right panel α = β = 0.4, we note that in the right panel that the curve is higher than the curve in the left panel. In Fig. (6) at $m = -1$ of P-states of bottomonium. Some states do not see since complex eigen energy appear by the negative sign of $m_l$. This is an agreement with Ref. [39]. Also, the generalized fractional plays role where in the left panel we take α = β = 0.9 and in the right panel α = β = 0.4, we note that in the right panel the curve is higher than the curve in the left panel.

**Table (1):** Mass Spectra of Charmonium (in GeV) for two cases (the first case in classical (α = β = 1) and two cases in fractional (α = 0.84, β = 1) at λ = 0.8)

| State | Present work of the First case | Present work of the second case | [34] | [35] | [36] | [37] | Exp.[40] | Error of state of first case | Error of state of the second case |
|---|---|---|---|---|---|---|---|---|---|
| 1S | 3.096 | 3.074 | 3.078 | 3.096 | 3.078 | 3.239 | 3.096 | 0 | 0.007 |
| 2S | 3.619 | 3.582 | 4.187 | 3.686 | 3.581 | 3.646 | 3.649 | 0.008 | 0.018 |
| 3S | 4.142 | 4.090 | 5.297 | 3.984 | 4.085 | 4.052 | 4.040 | 0.025 | 0.012 |
| 4S | 4.665 | 4.598 | 6.407 | 4.150 | 4.589 | 4.459 | 4.415 | 0.056 | 0.041 |
| 1P | 3.488 3.619 3.75 | 3.513 3.654 3.796 | 3.415 | 3.433 | 3.415 | 3.372 | 3.525 | 0.010 | 0.003 |
| 2P | 4.011 4.142 4.273 | 4.020 4.163 4.305 | 4.143 | 3.910 | 3.917 | 3.779 | 3.900 | 0.005 | 0.007 |
| 1D | 3.358 3.75 3.142 4.534 4.927 | 5.067 4.644 4.221 3.796 3.369 | 3.752 | 3.767 | 3.749 | 3.604 | 3.769 | 0.028 | 0.030 |
| Total Error | 0.018 | 0.016 | 0.9513 | 0.11065 | 0.11152 | 0.05788 | - | | |

**Table (2):** Mass Spectra of Charmonium (in GeV) for two cases (the first case in classical ($\alpha = \beta = 1$) at $\lambda = 0.7$ and two cases in fractional ($\alpha = 0.8, \beta = 0.9$) at $\lambda = 0.8$))

| State | Present work of the first case | Present work of the second case | [34] | [36] | [38] | Exp.[41] | Error of state of the first case | Error of state of the second case |
|---|---|---|---|---|---|---|---|---|
| 1S | 9.460 | 9.465 | 9.510 | 9.510 | 9.460 | 9.460 | 0 | 0.0005 |
| 2S | 9.844 | 9.853 | 10.627 | 10.038 | 10.023 | 10.023 | 0.018 | 0.016 |
| 3S | 10.288 | 10.241 | 11.726 | 10.566 | 10.585 | 10.355 | 0.006 | 0.011 |
| 4S | 10.611 | 10.628 | 12.834 | 11.094 | 11.148 | 10.580 | 0.003 | 0.004 |
| 1P | 9.679 9.843 10.008 | 9.802 9.911 10.020 | 9.862 | 9.862 | 9.492 | 9.900 | 0.006 | 0.001 |
| 2P | 10.063 10.228 10.392 | 10.19 10.299 10.408 | 10.944 | 10.390 | 10.038 | 10.260 | 0.003 | 0.003 |
| 1D | 9.898 10.063 10.227 10.392 10.557 | 10.129 10.237 10.345 110.733 11.120 | 10.214 | 10.214 | 9.551 | 10.161 | 0.006 | 0.003 |
| Total error | 0.006 | 0.0055 | 0.484 | 0.0137 | 0.028 | - | | |

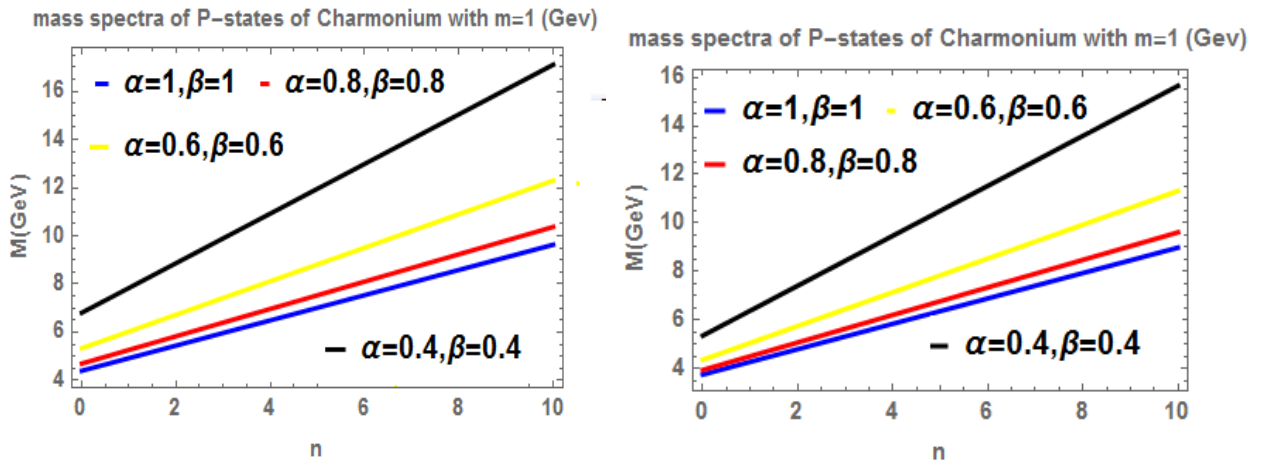

**Fig. (1):** *(Left panel): Mass spectra (in GeV) of Charmonium at various values of $\alpha, \beta$ of the quantum number n at topological defect ($\lambda = 0.4$) at m=+1. (Right panel): Mass spectra (in GeV) of Charmonium at various values $\alpha, \beta$ of n at topological defect ($\lambda = 0.8$) at m=+1.*

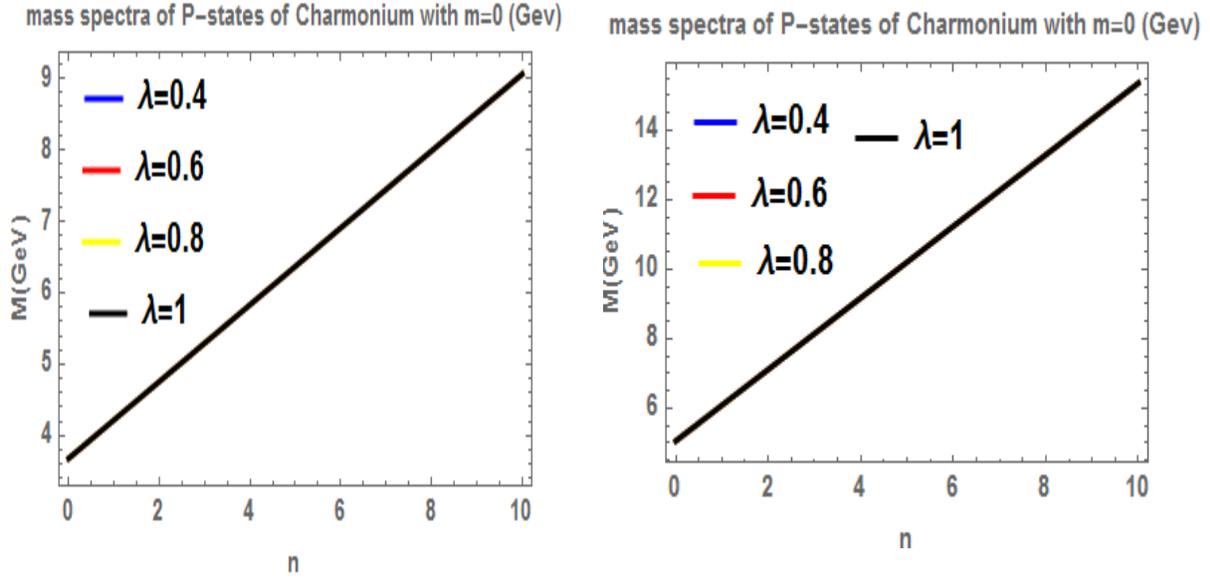

**Fig. (2):** *(Left panel): Mass spectra (in GeV) of Charmonium of the n at various values of topological defect λ at (α = β = 0.9) at m=0. (Right panel): Mass spectra (in GeV) of Charmonium of the n at various values of topological defect λ at (α = β = 0.4. ) at m=0.*

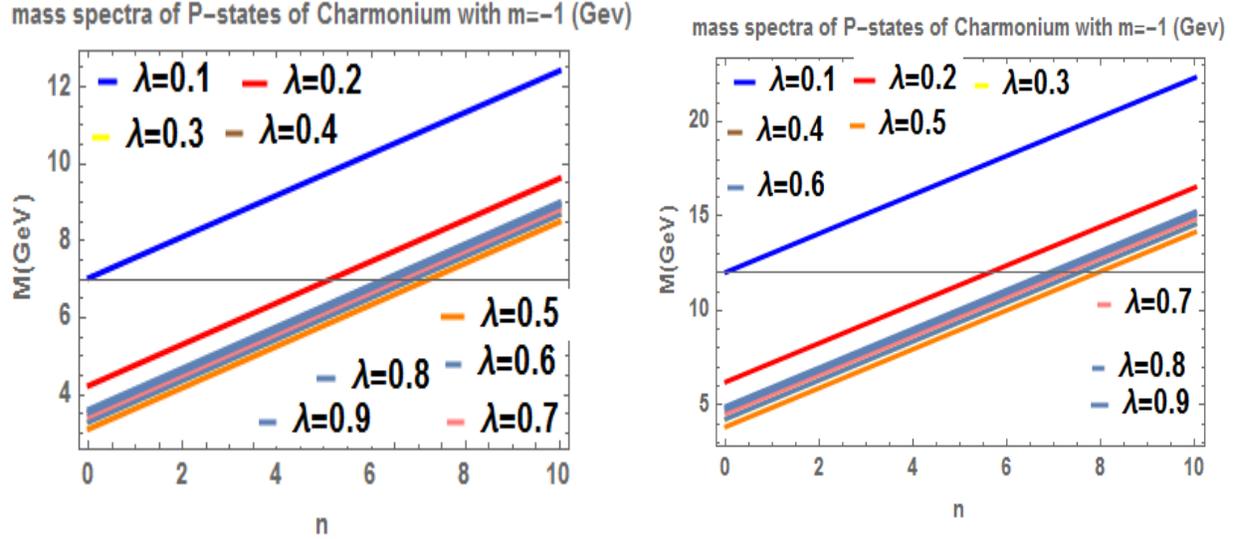

**Fig. 3:** *(Left panel): Mass spectra (in GeV) of Charmonium of the quantum number n at different values of topological defect λ at α = β = 0.9. )at m=-1.( Right panel):Mass spectra (in GeV) of Charmonium of the quantum number n at different values of topological defect λ at α = β = 0.4. ) at m=-1.*

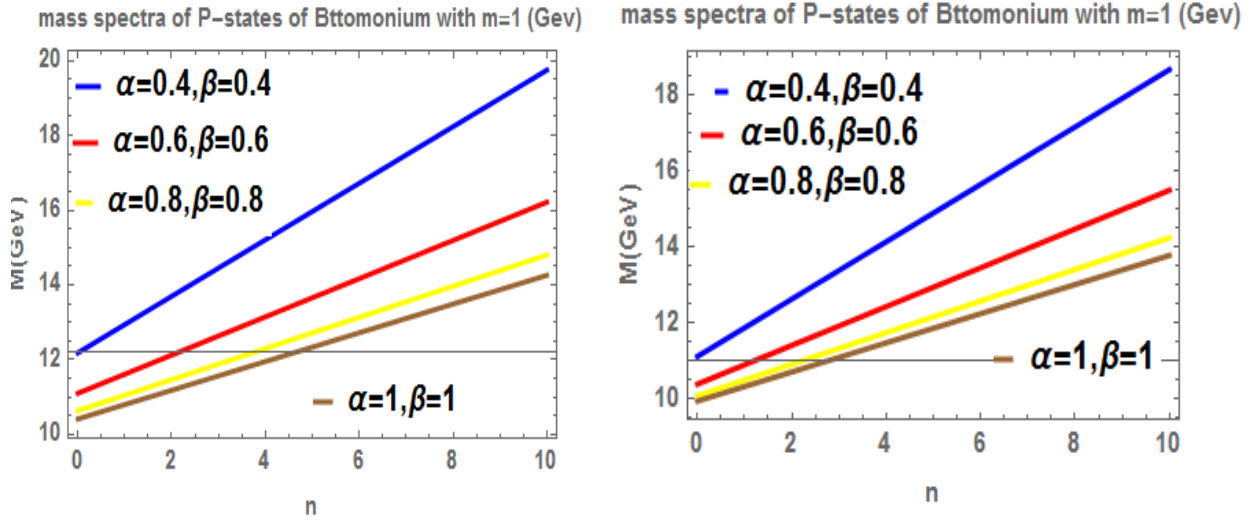

**Fig. (4):** *(Left panel): Mass spectra (in GeV) of Bottomonium of the n at various values of α, β at topological defect (λ = 0.4) at m=+1. (Right panel): mass spectra (in GeV) of Bottomonium of the n at various values of α, β at topological defect (λ = 0.8) at m=+1.*

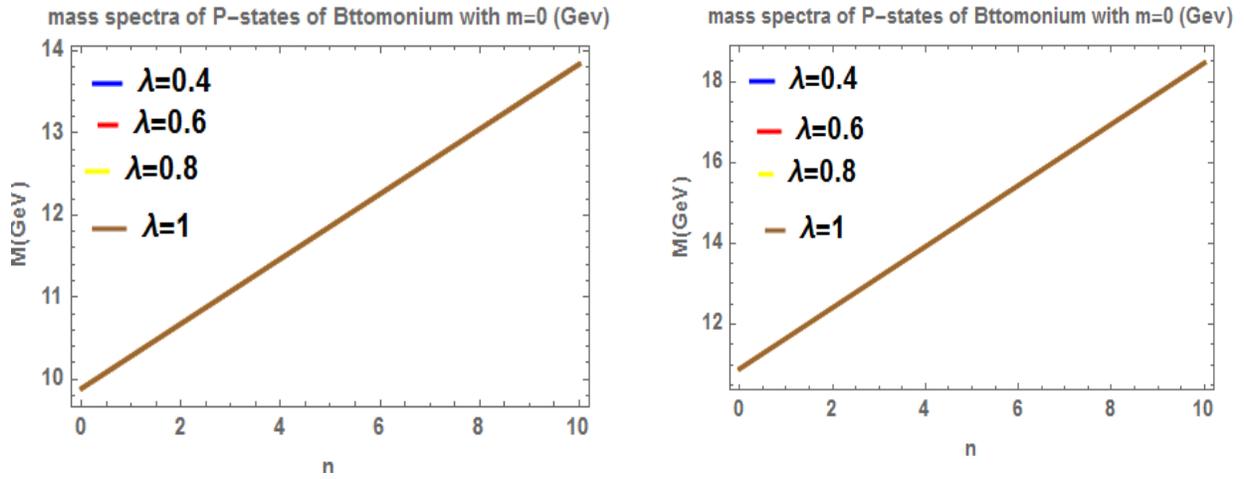

**Fig. (5):** *(Left panel): Mass spectra (in GeV) of Bottomonium of the n at various values of topological defect λ at (α = β = 0.9.) at m=0. (Right panel): Mass spectra (in GeV) of Bottomonium of the n at various values of topological defect λ at (α = β = 0.4.) at m=0.*

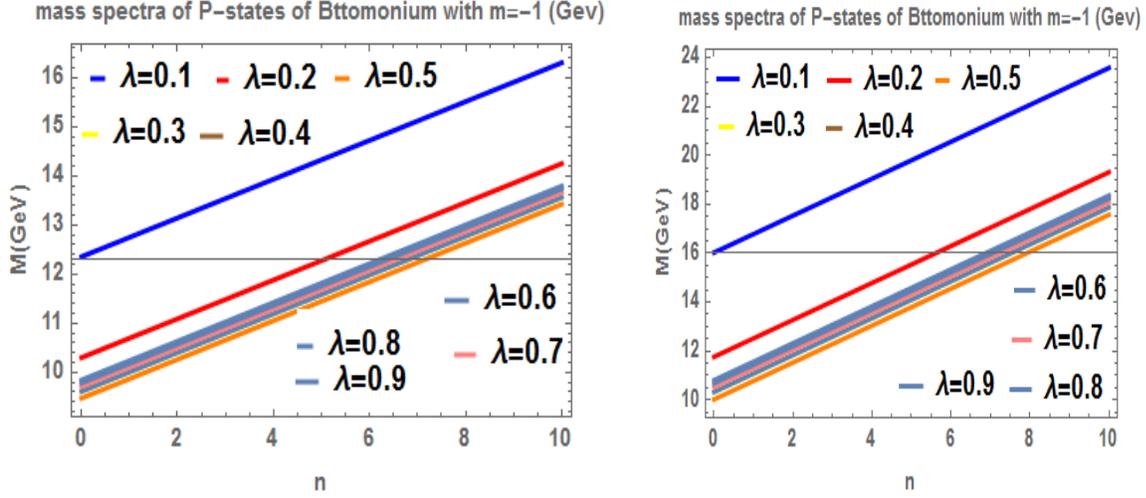

**Fig. (6):** *(Left panel): Mass spectra (in GeV) of Bottomnium of the n at various values of topological defect λ at α = β = 0.9. ) at m=-1. (Right panel) Mass spectra (in GeV) of Bottomonium of the n at various values of topological defect λ at (α = β = 0.4. ) at m=-1.*

### 4.3. The radial expectation values

The radial expectation values were given by the relation

$$\langle y^g \rangle = \int_0^\infty y^g \ |\psi_{nl}(y)|^2 dy \tag{43}$$

Table. (3), the radial mean value $\langle y \rangle$ $(GeV)^{-1}$ for various levels of Charmonium and topological defect causes to split n P and n D states. The mean value were calculated with the normalized wave function in two cases (first case ( $\alpha = \beta = 1$, $\lambda = 0.1$)). The result in this case is a good with other works but in the second case ($\alpha = 0.6, \beta = 0.27, \ \lambda = 0.9$) the generalized fractional parameter plays an important role where the results is close with other works [42, 43, 44, 45].

Table (4), the radial mean value $\langle y^{-1} \rangle$ (GeV) for various levels of Charmonium and topological defect causes to split n P and n D states. The mean value were calculated with the normalized wave function in two cases (first case ( $\alpha = \beta = 1$, $\lambda = 0.15$)) the results in this case is a good with other works but in the second case ( $\alpha = 0.59, \beta = 0.2, \lambda = 0.6$) the generalized fractional parameter plays an important role where the results is close with other works [42, 43, 44, 45]. In Table (5), the radial mean value $\langle y \rangle$ $(GeV)^{-1}$ for various levels of bottomonium and topological defect causes to split nP and nD states. The mean value were calculated using the normalized wave function in two cases (first case ( $\alpha = \beta = 1$, $\lambda = 0.13$)), we note that the effect topological defect on S-states does not appear like

P-states and D-states, the results in this case is a good with other works but in the second case ( $\alpha = 0.7, \beta = 0.1, \lambda = 0.6$), the generalized fractional prameter plays an important role and we can see this specially in S-states and P-states and D-states where the results is close with other works [42, 43, 44]. Also, we can see that the average value $\langle y \rangle$ $(GeV)^{-1}$ decreases for bottomonium with an increase in the reduced mass compared with charmonium. In Table. (6), the radial mean value $\langle y^{-1} \rangle$ (GeV) for various levels of bottomonium and topological defect causes to split nP and nD states. The mean value were calculated with the normalized wave function in two cases (first case ( $\alpha = \beta = 1, \lambda = 0.13$)), we note that the effect topological defect on S-states does not appear like P-states and D-states, the results in this case is a good with other works [42, 43, 44] but in the second case ( $\alpha = 0.7, \beta = 0.1, \lambda = 0.6$), the generalized fractional parameter plays an important role and we can see this specially in S-states and P-states and D-states where the results increase for bottomonium with an increase in the reduced mass compared with charmonium.

**Table (3)**. Expectation value of Charmonium of $\langle y \rangle$ $(GeV)^{-1}$ for two cases (the first case ( $\alpha = \beta = 1, \lambda = 0.1$), the second case ( $\alpha = 0.6, \beta = 0.27, \lambda = 0.9$).

| State | Present of first case $\langle y \rangle$ $(GeV)^{-1}$ | Present of second case $\langle y \rangle (GeV)^{-1}$ | [42] | [42] | [43] | [44] | [45] |
|---|---|---|---|---|---|---|---|
| 1S | 1.184 | 3.030 | 3.073 | 3.086 | 2.619 | 2.790 | 1.002 |
| 2S | 1.601 | 4.732 | 5.770 | 5.777 | 4.761 | 4.612 | 1.551 |
| 1P | 4.170<br>1.545<br>4.722 | 3.554<br>3.664<br>3.779 | 4.331 | 4.325 | - | 4.266 | - |
| 2P | 4.347<br>1.894<br>4.883 | 5.231<br>5.337<br>5.447 | 7.511 | 7.488 | 3.725 | 5.588 | - |
| 1D | 6.137<br>3.873<br>2.025<br>4.980<br>6.945 | 4.529<br>4.660<br>4.792<br>4.925<br>5.059 | 6.410 | 6.380 | | | - |

**Table. (4).** Expectation value of Charmonium of $\langle y^{-1} \rangle$ (GeV) for two cases (the first case ($\alpha = \beta = 1, \lambda = 0.15$), the second case ($\alpha = 0.59, \beta = 0.2, \lambda = 0.6$).

| State | Present of first case $\langle y^{-1} \rangle$ (GeV) | Present of second case $\langle y^{-1} \rangle$ (GeV) | [42] | [42] | [43] | [44] |
|---|---|---|---|---|---|---|
| 1S | 1.173 | 0.502 | 0.456 | 0.454 | 0.492 | 0.507 |
| 2S | 0.805 | 0.268 | 0.251 | 0.250 | 0.325 | 0.396 |
| 1P | 0.342<br>0.795<br>0.276 | 0.308<br>0.393<br>0.485 | 0.312 | 0.312 | - | 0.316 |
| 2P | 0.319<br>0.628<br>0.263 | 0.197<br>0.231<br>0.262 | 0.188 | 0.188 | 0.307 | 0.239 |
| 1D | 0.226<br>0.396<br>0.562<br>0.254<br>0.184 | 0.442<br>0.347<br>0.275<br>0.224<br>0.189 | 0.202 | 0.203 | - | |

**Table (5)** Expectation value of Bottomonium of $\langle y \rangle$ $(GeV)^{-1}$ for two cases (the first case ($\alpha = \beta = 1, \lambda = 0.13$), the second case ($\alpha = 0.7, \beta = 0.1, \lambda = 0.6$).

| State | Present of first case $\langle y \rangle$ $(GeV)^{-1}$ | Present of second case $\langle y \rangle$ $(GeV)^{-1}$ | [42] | [42] | [43] | [44] |
|---|---|---|---|---|---|---|
| 1S | 0.600 | 1.874 | 1.615 | 1.556 | 1.823 | 1.574 |
| 2S | 0.819 | 2.785 | 2.887 | 2.736 | 3.100 | 2.523 |
| 1P | 1.861<br>0.794<br>2.218 | 1.888<br>1.983<br>2.140 | 2.105 | 1.981 | - | 2.306 |
| 2P | 1.974<br>0.980<br>2.318 | 2.796<br>2.880<br>3.019 | 3.551 | 3.309 | 2.446 | 3.017 |
| 1D | 2.790<br>1.666<br>1.058 | 1.925<br>2.056<br>2.231 | 2.956 | 2.732 | - | - |

| | 2.382 | 2.428 | | | | |
| | 3.226 | 2.634 | | | | |

**Table. (6)** Expectation value of Bottomonium of $\langle y^{-1}\rangle$ (GeV) for two cases (the first case ($\alpha = \beta = 1, \lambda = 0.2$), the second case ($\alpha = 0.56, \beta = 0.1, \lambda = 0.5$).

| State | Present of first case $\langle y^{-1}\rangle$ (GeV) | Present of first case $\langle y^{-1}\rangle$ (GeV) | [42] | [42] | [43] | [44] |
|---|---|---|---|---|---|---|
| 1S | 2.352 | 0.847 | 0.848 | 0.874 | 0.686 | 0.8706 |
| 2S | 1.593 | 0.478 | 0.493 | 0.518 | 0.486 | 0.6946 |
| 1P | 0.847<br>1.565<br>0.614 | 0.844<br>0.770<br>0.658 | 0.632 | 0.668 | - | 0.467 |
| 2P | 0.759<br>1.226<br>0.573 | 0.447<br>0.452<br>0.410 | 0.393 | 0.420 | 0.467 | 0.4421 |
| 1D | 0.548<br>1.083<br>1.083<br>0.548<br>0.398 | 0.835<br>0.745<br>0.634<br>0.538<br>0.464 | 0.435 | 0.469 | | |

## 4.4. Thermodynamic properties of heavy quarkonia.

The partition function serves as the foundation for thinking about the thermodynamics properties of heavy quarkonia inside the cosmic-string framework [46, 47]. According to statistical mechanics, the partition function can be built as follows:

### 4.4.1. Partition function

$$Z(\beta_s) = \sum_{n=0}^{\infty} e^{-\beta_s E_n} =$$

$$\frac{1}{2} e^{-\beta_s \left[d - \frac{b^2}{4a} + \zeta^2 \sqrt{\frac{2a}{M}} \left[\alpha + \sqrt{\frac{\alpha^2}{4} + \frac{1}{\zeta^2}\left[l - \left(1 - \frac{1}{\lambda}\right)m\right]\left[l - \left(1 - \frac{1}{\lambda}\right)m + 1\right]}\right]\right]} \text{Csch}\left[\frac{\alpha}{2}\beta_s \zeta^2 \sqrt{\frac{2a}{M}}\right] \qquad (44)$$

where $\beta_s = \frac{1}{KT}$, the Boltzmann constant is $K$, and the system's absolute temperature is T.

In Figs. (7, 8), we plot the partition function of P-states of charmonium and bottomonium as function of $\beta_s$ at different values from $\alpha, \beta$ under effect topological defect. We note that by increasing the values of $\alpha, \beta$ the curve becomes higher. In left panel, we take topological defect $\lambda = 0.2$, we note the curves are not connected to the standard Minkowski curve, we note that also the splitting increase as $\lambda$ decrease and $Z(\beta_s)$ decreases with increasing $\beta_s$ this is in agreement with [48, 49, 50]. The author of [48] within the framework of the NU approach, we have used DFDEP to solve the Klein-Gordon equation, they obtained the energy eigenvalues and associated wave function in D dimensions in great detail and noticed that the $Z(\beta)$ for the Minkowskian case monotonically lowers with increasing. In Ref. [49], the partition function Z decreases monotonically with increasing λ for the two diatomic molecules considered, and reaches a constant value for some typical values of λ. In Ref. [50], the author plotted the partition function of P-states of charmonium and bottomonium as function of $\beta_s$ at different values topological defect and they show that by increasing topological defect $Z(\beta)$ decreases with increasing $\beta$. In Figs. (9, 10), we plot the free energy of P-states of charmonium and bottomonium as function of $\beta_s$ at different values from $\alpha, \beta$ under effect topological defect. We note that by increasing the values of $\alpha, \beta$ the curve becomes higher. In the left panel, we take topological defect $\lambda = 0.2$, we note The curves are not connected to the standard Minkowski curve. We note that also the splitting increase as $\lambda$ decrease and free energy increases with increasing $\beta_s$ and we note that free energy $(F)$ decreases by increasing temperature. In Ref. [51], they show $F$ decreases by increasing temperature of neutral particle. This is in agreement with [46, 47, 50, 51]. In Figs. (11, 12), we plot the mean energy of P-states of charmonium and bottomonium as function of $\beta_s$ at different values from $\alpha, \beta$ under effect topological defect. We note that by increasing the values of $\alpha, \beta$ the curve becomes lower. In the left panel, we take topological defect $\lambda = 0.2$, we note the curves are separated from the classical Minkowski curve we note that also the splitting increase as $\lambda$ decrease and mean energy $(U)$ decreases with increasing $\beta_s$ and by increasing the parameter of topological defect its values shift to lower values. In Ref. [51], they calculated thermodynamic properties of neutral particle under effect topological defect from non-relativistic Schrödinger-Pauli equation. The author noted that by increasing temperature the $(U)$ increases. This is an agreement with [49, 49, 50, 51]. In Figs. (13, 14), we plot the specific heat $(C)$ of P-states of charmonium and bottomonium as a function of $\beta_s$ at different values from topological defect under effect $\alpha, \beta$. We note that by increasing topological defect specific heat decreases, this is agreement with [52], and also in right panel we take $\alpha = \beta = 0.1$ we note that by increasing $\alpha, \beta$ the curves became lower as in the left panel.

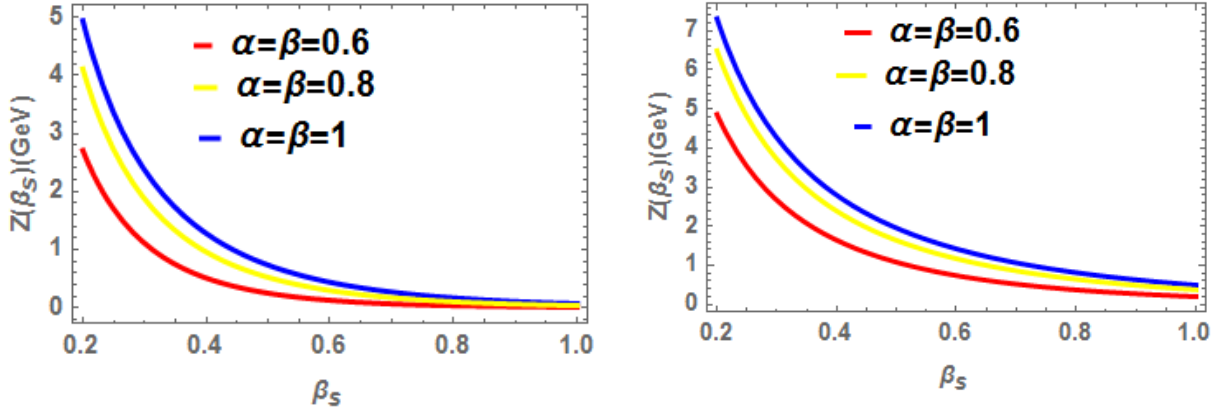

**Fig. (7)** (Left panel): The partition function (Z) for second excited states of $c\bar{c}$ is shown as a function of $\beta_s$, for different values of $\alpha$ and $\beta$ with $l = 1$, $m_l = +1$ at topological defect ($\lambda = 0.2$). (Right panel): The partition function (Z) for second excited states of $c\bar{c}$ is shown as a function of $\beta_s$, for different values of $\alpha$ and $\beta$ with $l = 1$, $m_l = +1$ at topological defect ($\lambda = 0.8$).

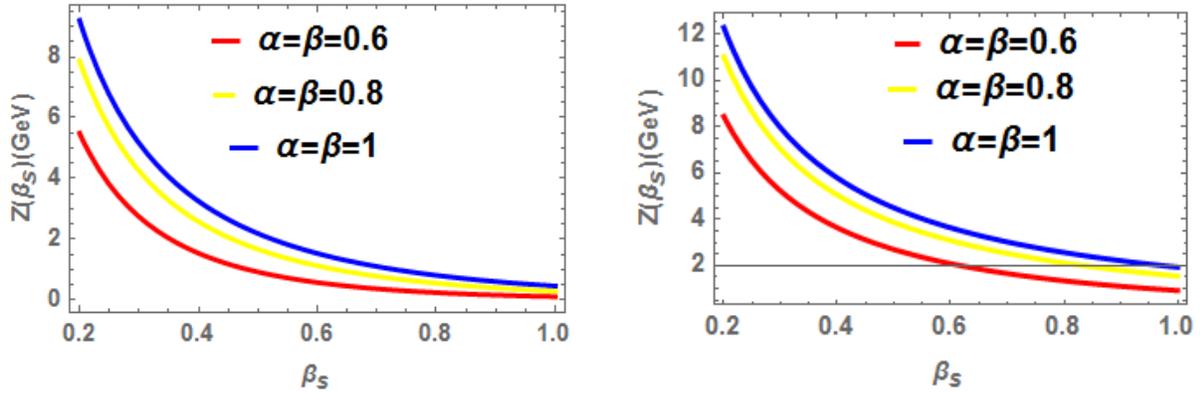

**Fig. (8)** (Left panel): The partition function (Z) for second excited states of $b\bar{b}$ is shown as a function of $\beta_s$, for different values of $\alpha$ and $\beta$ with $l = 1$, $m_l = +1$ at topological defect ($\lambda = 0.2$). (Right panel): The partition function (Z) for second excited states of $c\bar{c}$ is shown as a function of $\beta_s$, for different values of $\alpha$ and $\beta$ with $l = 1$, $m_l = +1$ at topological defect ($\lambda = 0.8$).

## 4.3.2 Free energy F

$$F(\beta_s) = -\frac{1}{\beta_s} lnZ(\beta_s) \tag{45}$$

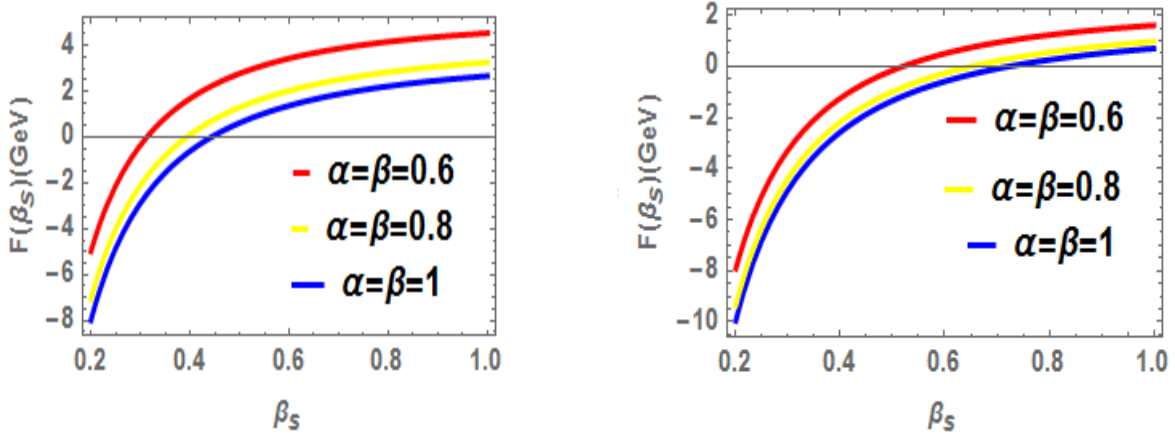

**Fig. (9)** (Left panel): The free energy is plotted as for second excited states of $c\bar{c}$ is shown as a function of $\beta_s$, for different values of $\alpha$ and $\beta$ with $l = 1$, $m_l = +1$ at topological defect ($\lambda = 0.2$). (Right panel): The free energy for second excited states of $c\bar{c}$ is shown as a function of $\beta_s$, for different values of $\alpha$ and $\beta$ with $l = 1$, $m_l = +1$ at topological defect ($\lambda = 0.8$).

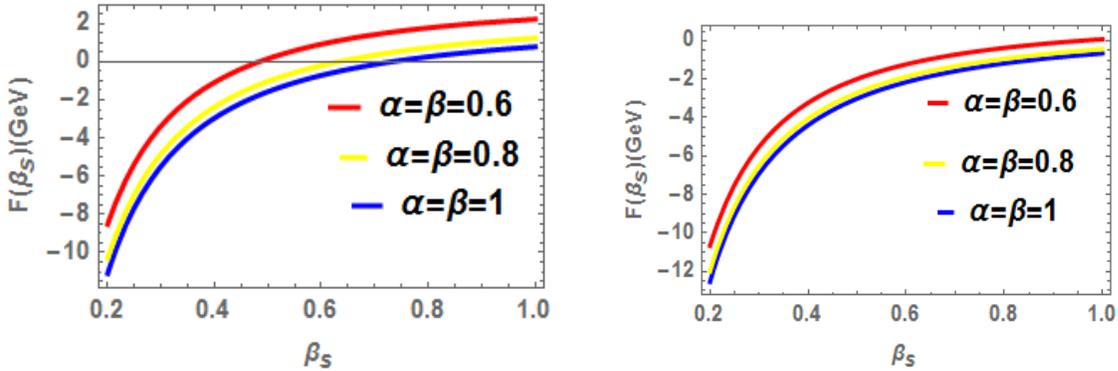

**Fig. (10)** (Left panel): The free energy is plotted as for second excited states of $b\bar{b}$ is shown as a function of $\beta_s$, for different values of $\alpha$ and $\beta$ with $l = 1$, $m_l = +1$ at topological defect ($\lambda = 0.2$). (Right panel): The free energy for second excited states of $c\bar{c}$ is shown as a function of $\beta_s$, for different values of $\alpha$ and $\beta$ with $l = 1$, $m_l = +1$ at topological defect ($\lambda = 0.8$).

### 4.4.3 Mean energy U

$$U(\beta_s) = -\frac{\partial}{\partial(\beta_s)} lnZ(\beta_s) \tag{46}$$

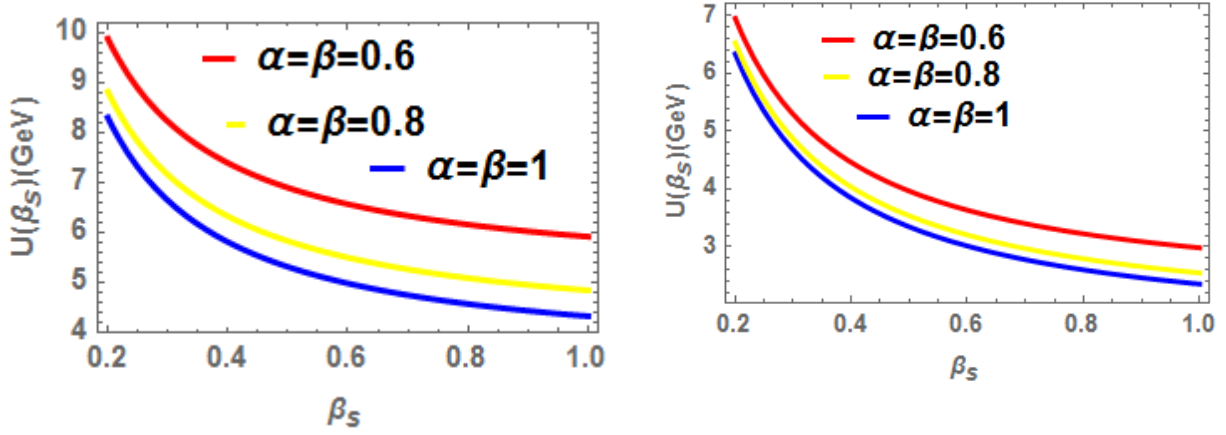

**Fig. (11).** (Left panel): The mean energy is plotted as for second excited states of $c\bar{c}$ is shown as a function of $\beta_s$, for different values of $\alpha$ and $\beta$ with $l = 1$, $m_l = +1$ at topological defect ($\lambda = 0.2$). (Right panel): The mean energy for second excited states of $c\bar{c}$ is shown as a function of $\beta_s$, for different values of $\alpha$ and $\beta$ with $l = 1$, $m_l = +1$ at topological defect ($\lambda = 0.8$).

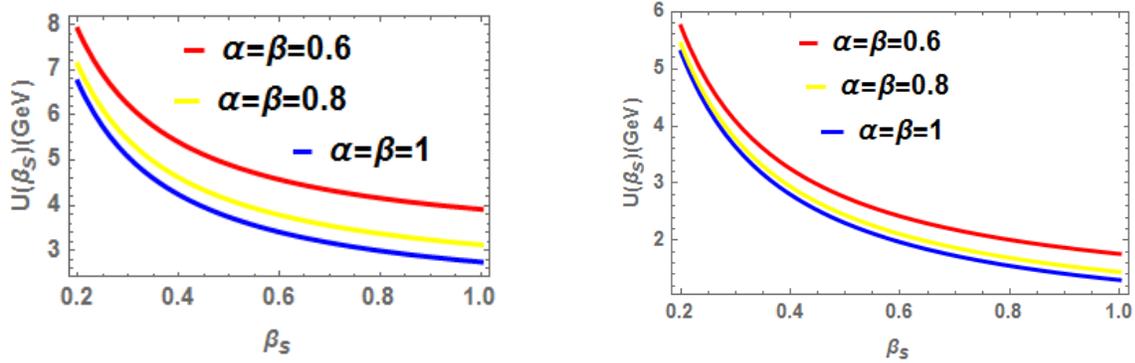

**Fig. (12).** (Left panel): The mean energy is plotted as for second excited states of $b\bar{b}$ is shown as a function of $\beta_s$, for different values of $\alpha$ and $\beta$ with $l = 1$, $m_l = +1$ at topological defect ($\lambda = 0.2$). (Right panel): The mean energy for second excited states of $b\bar{b}$ is shown as a function of $\beta_s$, for different values of $\alpha$ and $\beta$ with $l = 1$, $m_l = +1$ at topological defect ($\lambda = 0.8$).

### 4.4.4 Specific heat C

$$C(\beta_s) = \frac{\partial (U)}{\partial T} = -K\,\beta_s^{2}\,\frac{\partial (U)}{\partial (\beta_s)} \tag{47}$$

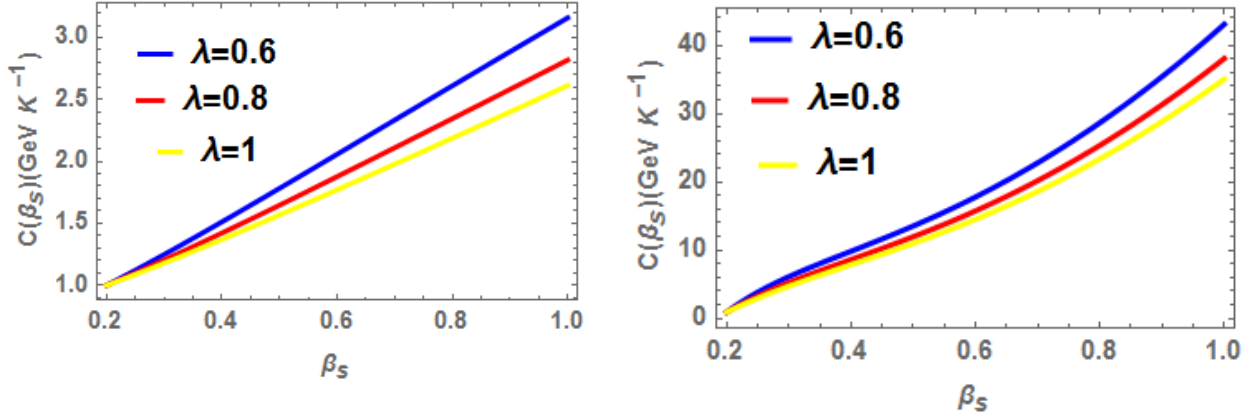

**Fig. (13).** (left panel): The specific heat is plotted as for second excited states of $c\bar{c}$ is shown as a function of $\beta_s$, for different values of topological defect at $\alpha = \beta = 0.8$ with $l = 1$, $m_l = +1$. (Right panel): The specific heat is plotted as for second excited states of $c\bar{c}$ is shown as a function of $\beta_s$, for different values of topological defect at $\alpha = \beta = 0.1$ with $l = 1$, $m_l = +1$.

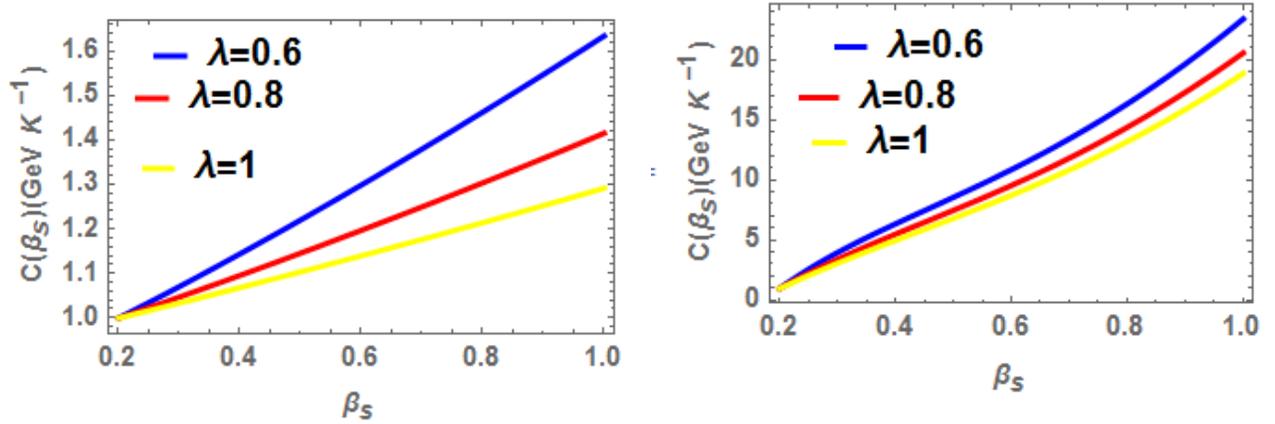

**Fig. (14).** (Left panel): The specific heat is plotted as for second excited states of $b\bar{b}$ is shown as a function of $\beta_s$, for different values of topological defect at $\alpha = \beta = 0.8$ with $l = 1$, $m_l = +1$. (Right panel): The specific heat is plotted as for second excited states of $b\bar{b}$ is shown as a function of $\beta_s$, for different values of topological defect at $\alpha = \beta = 0.1$ with $l = 1$, $m_l = +1$.

## 5. Conclusion

In this paper, we solved the radial Schrödinger equation in the a cosmic string with extended Cornell potential by using generalized fractional derivative of extended Nikorov-Uvarov (GFD-ENU) method under effect topological defect. We obtained eigen energy of heavy quarkonia and wave functions in the fractional form. Firstly, we obtained the special cases from the general case then we calculated the mass of charmonium and bottomonium in two cases (the first case is the classical case and second is the fractional case) under effect topological defect, it has been noted that the splitting between nP and nD states is caused by the presence of the topological defect. The excited states are split into $2l + 1$ components, suggesting that a topological defect's gravity field interacts with energy levels in a way that is comparable to the Zeeman effect brought on by the magnetic field and where the results are close with experimental data and good with other works such as Refs. [34, 35, 36, 37, 38]. We note that generalized fractional derivative plays an important role in this work since we obtain lower total error in calculating the mass of heavy quarkonia. We also obtained the wave function then we calculated the root mean of charmonium and bottomonium in two cases (the first case is classical and second is fractional cases) under effect topological defect and the results are an agreement with Refs. [42, 43, 44, 45]. The mass and thermodynamics properties were analyzed graphically. With respect to the classical limit, the thermodynamic quantities exhibit a shift; When the cosmic-string parameter is low, this fluctuation becomes more important. Free energy increases with increasing $\beta_s$ and we note that free energy decreases by increasing temperature, we note that also the splitting increase as λ decrease and this is an agreement with Refs. [46, 47, 50, 51] by increasing the values of α, β the curve becomes higher. The $Z(\beta_s)$ shifts to lower values as $\lambda$ decreases this is an agreement with Refs. [48, 49, 50] and by increasing the values of $\alpha, \beta$ the curve becomes higher values. The Minkowski curve is separated from the curves with λ≠1, Then, by raising the $\lambda$, the mean energy values are changed to lower values. this is an agreement with Refs. [48, 49, 50, 51] by increasing α, β the curves became lower. We note that by increasing topological defect specific heat decreases, this is agreement with [52], by increasing α, β the curves became lower.

# 6-References

1. Abu-Shady, M. "Chiral logarithmic quark model of N and Δ with an A-term in the mean-field approximation." International Journal of Modern Physics A 26.02 (2011): 235-249.
2. Abu-Shady, M., and E. M. Khokha. "Bound state solutions of the Dirac equation for the generalized Cornell potential model." International Journal of Modern Physics A 36.29 (2021): 2150195.
3. Rashdan, M., M. Abu-Shady, and T. S. T. Ali. "Extended linear sigma model in higher order mesonic interactions." International Journal of Modern Physics E 15.01 (2006): 143-152.
4. Abu-Shady, M., and H. M. Mansour. "Quantized linear." Physical Review C: Nuclear Physics 85.5 (2012).
5. Abu-Shady, M., and Sh Y. Ezz-Alarab. "Conformable fractional of the analytical exact iteration method for heavy quarkonium masses spectra." Few-Body Systems 62 (2021): 1-8.
6. Abu-Shady, M. "The effect of finite temperature on the nucleon properties in the extended linear sigma model." International Journal of Modern Physics E 21.06 (2012): 1250061.
7. Abu-Shady, M., and M. Soleiman. "The extended quark sigma model at finite temperature and baryonic chemical potential." Physics of Particles and Nuclei Letters 10 (2013): 683-692.
8. Abu-Shady, M., H. M. Mansour, and A. I. Ahmadov. "Dissociation of quarkonium in hot and dense media in an anisotropic plasma in the nonrelativistic quark model." Advances in High Energy Physics 2019 (2019).
9. Kibble, Thomas WB. "Topology of cosmic domains and strings." Journal of Physics A: Mathematical and General 9.8 (1976): 1387.
10. Vilenkin, Alexander, and E. Paul S. Shellard. Cosmic strings and other topological defects. Cambridge University Press, 1994.
11. Davis, A. C., and T. W. B. Kibble. "Fundamental cosmic strings." Contemporary Physics 46.5 (2005): 313-322.
12. Kibble, Thomas Walter Bannerman, George Lazarides, and Qaisar Shafi. "Walls bounded by strings." Physical Review D 26.2 (1982): 435
13. Rocher, Jonathan. Contraintes cosmologiques sur la physique de l'univers primordial. Diss. Université Paris Sud-Paris XI, 2005.

14. Audretsch, Jürgen, and Athanasios Economou. "Conical bremsstrahlung in a cosmic-string spacetime." Physical Review D 44.12 (1991): 3774
15. Harari, Diego D., and Vladimir D. Skarzhinsky. "Pair production in the gravitational field of a cosmic string." Physics Letters B 240.3-4 (1990): 322-326.
16. Florkowski, Wojciech. Phenomenology of ultra-relativistic heavy-ion collisions. World Scientific Publishing Company, 2010
17. Kakade, Uttam, and Binoy Krishna Patra. "Quarkonium dissociation at finite chemical potential." Physical Review C 92.2 (2015): 024901
18. Matsui, Tetsuo, and Helmut Satz. "J/ψ suppression by quark-gluon plasma formation." Physics Letters B 178.4 (1986): 416-422
19. Filinov, V. S., et al. "Thermodynamics of the Quark-Gluon Plasma at Finite Chemical Potential: Color Path Integral Monte Carlo Results." Contributions to Plasma Physics 55.2-3 (2015): 203-208.
20. Schleif, M., and R. Wünsch. "Thermodynamic properties of the SU (2) f chiral quark–loop soliton." The European Physical Journal A-Hadrons and Nuclei 1 (1998): 171-186.
21. Abu-Shady, M. "Meson properties at finite temperature in the linear sigma model." International Journal of Theoretical Physics 49 (2010): 2425-2436.
22. Nikiforov, Arnold F., and Vasiliĭ Borisovich Uvarov. Special functions of mathematical physics. Vol. 205. Basel: Birkhäuser, 1988.
23. Karayer, Hale, Doğan Demirhan, and F. Büyükkılıç. "Extension of Nikiforov-Uvarov method for the solution of Heun equation." Journal of Mathematical Physics 56.6 (2015).
24. Abu-Shady, M. "Quarkonium masses in a hot QCD medium using conformable fractional of the Nikiforov–Uvarov method." International Journal of Modern Physics A 34.31 (2019): 1950201..
25. Abu-Shady, M., E. M. Khokha, and T. A. Abdel-Karim. "The generalized fractional NU method for the diatomic molecules in the Deng–Fan model." The European Physical Journal D 76.9 (2022): 159.
26. Abu-Shady, M., and Mohammed KA Kaabar. "A generalized definition of the fractional derivative with applications." Mathematical Problems in Engineering 2021 (2021): 1-9.
27. Katanaev, M. O., and I. V. Volovich. "Theory of defects in solids and three-dimensional gravity." Annals of Physics 216.1 (1992): 1-28.
28. Furtado, Claudio, and Fernando Moraes. "On the binding of electrons and holes to disclinations." Physics Letters A 188.4-6 (1994): 394-396.

29. Kumar, Brijesh, M. B. Paranjape, and U. A. Yajnik. "Fate of the false monopoles: Induced vacuum decay." Physical Review D 82.2 (2010): 025022.
30. Lee, Bum-Hoon, et al. "Battle of the bulge: Decay of the thin, false cosmic string." Physical Review D 88.10 (2013): 105008.
31. Bergström, Lars, and Ariel Goobar. Cosmology and particle astrophysics. Springer Science & Business Media, 2006.
32. Inyang, E. P., et al. "Analytic study of thermal properties and masses of heavy mesons with quarkonium potential." Results in Physics 39 (2022): 105754.
33. Atangana Likéné, A., et al. "Nonrelativistic quark model for mass spectra and decay constants of heavy-light mesons using conformable fractional derivative and asymptotic iteration method." International Journal of Modern Physics A 37.35 (2022): 2250229.
34. Kumar, Ramesh, and Fakir Chand. "Asymptotic study to the N-dimensional radial Schrödinger equation for the quark-antiquark system." Communications in Theoretical Physics 59.5 (2013): 528.
35. Maksimenko, N. V., and S. M. Kuchin. "Determination of the mass spectrum of quarkonia by the Nikiforov–Uvarov method." Russian Physics Journal 54.1 (2011): 57-65.
36. Kumar, Ramesh, and Fakir Chand. "Series solutions to the N-dimensional radial Schrödinger equation for the quark–antiquark interaction potential." Physica Scripta 85.5 (2012): 055008.
37. Abu-Shady, M., and Sh Y. Ezz-Alarab. "Trigonometric Rosen–Morse potential as a quark–antiquark interaction potential for meson properties in the non-relativistic quark model using EAIM." Few-body systems 60.4 (2019): 66.
38. Rani, Richa, S. B. Bhardwaj, and Fakir Chand. "Mass spectra of heavy and light mesons using asymptotic iteration method." Communications in Theoretical Physics 70.2 (2018): 179.
39. Likéné, André Aimé Atangana, et al. "Effects of Gravitational Field of a Topological Defect on Heavy Quarkonia Spectra in a Non-Relativistic Quark Model." (2023).
40. M. Tanabashi et al. (Particle Data Group). "Review of Particle Physics". Physical Review D. 98 (3): 030001, (2018)
41. R. M. Barnnett et al, (Particle Data Group). "Review of Particle Physics". Physical Review D. 541 (1996).

42. Omugbe, E., et al. "Approximate mass spectra and root mean square radii of quarkonia using Cornell potential plus spin-spin interactions." Nuclear Physics A 1034 (2023): 122653.
43. Boroun, G. R., and H. Abdolmalki. "Variational and exact solutions of the wavefunction at origin (WFO) for heavy quarkonium by using a global potential." Physica Scripta 80.6 (2009): 065003.
44. Gupta, Pramila, and Indira Mehrotra. "Study of heavy quarkonium with energy dependent Potential." Journal of Modern Physics 3.10 (2012): 1530.
45. Ramírez Zaldívar, D. A., F. Guzmán, and D. Arrebato. "Charmonio: comparación entre modelos potenciales." Nucleus 63 (2018): 9-11.
46. Modarres, Majid, and Ahmad Mohamadnejad. "The thermodynamic properties of weakly interacting quark-gluon plasma via the one-gluon exchange interaction." Physics of Particles and Nuclei Letters 10 (2013): 99-104.
47. Modarres, M., and H. Gholizade. "Strange quark matter in the framework of one gluon exchange and density and temperature dependent particle mass models." International Journal of Modern Physics E 17.07 (2008): 1335-1355.
48. Ikot, A. N., et al. "Klein-Gordon equation particles in exponential-type molecule potentials and their thermodynamic properties in D dimensions." The European Physical Journal Plus 131 (2016): 1-17.
49. Yahya, W. A., and K. J. Oyewumi. "Thermodynamic properties and approximate solutions of the ℓ-state Pöschl–Teller-type potential." Journal of the Association of Arab Universities for Basic and Applied Sciences 21 (2016): 53-58.
50. Likéné, André, et al. "Effects of Gravitational Field of a Topological Defect on Statistical Properties of Heavy Quark-Antiquark Systems." East European Journal of Physics 3 (2022): 129-141.
51. Hassanabadi, Hassan, and Mansoureh Hosseinpour. "Thermodynamic properties of neutral particle in the presence of topological defects in magnetic cosmic string background." The European Physical Journal C 76 (2016): 1-7.
52. Edet, C. O., et al. "Magneto-transport and thermal properties of the Yukawa potential in cosmic string space-time." *Results in Physics* 39 (2022): 105749.